\documentclass[preprints,article,accept,pdftex,moreauthors]{Definitions/mdpi} 
\firstpage{1} 
\pubvolume{1}
\issuenum{1}
\articlenumber{0}
\pubyear{2025}
\copyrightyear{2025}
\datereceived{ } 
\daterevised{ } % Comment out if no revised date
\dateaccepted{ } 
\datepublished{ } 
\hreflink{https://doi.org/} % If needed use \linebreak
\Title{Cosmic magnification on high-redshift submillimeter galaxies}

\TitleCitation{Cosmic magnification on high-redshift SMGs}

\Author{Marcos M. Cueli $^{1,2,}$*\orcidA{}, Joaquín González-Nuevo $^{3,4}$\orcidB{}, Laura Bonavera $^{3,4}$\orcidC{} and Andrea Lapi $^{1,2,5,6}$\orcidD{}}

\AuthorNames{Marcos M. Cueli, et al.}

\isAPAStyle{%
       \AuthorCitation{Lastname, F., Lastname, F., \& Lastname, F.}
         }{%
        \isChicagoStyle{%
        \AuthorCitation{Lastname, Firstname, Firstname Lastname, and Firstname Lastname.}
        }{
        \AuthorCitation{Lastname, F.; Lastname, F.; Lastname, F.}
        }
}

\address{%
$^{1}$ \quad SISSA - Scuola Internazionale Superiore di Studi Avanzati, Via Bonomea 265, 34136, Trieste, Italy\\
$^{2}$ \quad IFPU - Institute of Fundamental Physics of the Universe, Via Beirut 2, 34014, Trieste, Italy\\
$^{3}$ \quad Departamento de Física, Universidad de Oviedo, Calle Federico García Lorca 18, 33007, Oviedo, Spain\\
$^{4}$ \quad ICTEA - Instituto de Ciencias y Tecnologías Espaciales de Asturias, Calle Independencia 13, 33004, Oviedo, Spain\\
$^{5}$ \quad IRA-INAF, Via Piero Gobetti 101, I–40129 Bologna, Italy\\
$^{6}$ \quad INFN - Istituto Nazionale Fisica Nucleare, Sezione di Trieste, Via Valerio 2, 34127 Trieste, Italy
}

\corres{Correspondence: mmunizcu@sissa.it}

\abstract{Weak lensing magnification probes the correlation between galaxies and the underlying matter field in a similar fashion to galaxy-galaxy lensing shear. Although it has long been sidelined in favor of the latter on the grounds of a poorer performance in terms of statistical significance, the provision of a large sample of high-redshift submillimeter galaxies by the \emph{Herschel} observatory has transformed the landscape of cosmic magnification due to their optimal physical properties for magnification analyses. This review aims to summarize the core principles and unique advantages of cosmic magnification on high-redshift submillimeter galaxies and discuss recent results applied for cosmological inference. The outlook and challenges of this observable are also outlined, with a focus on the ample scope for exploration and its potential to emerge as a competitive  independent cosmological probe.}


\begin{document}

%%%%%%%%%%%%%%%%%%%%%%%%%%%%%%%%%%%%%%%%%%

\section{Introduction}

Over the past decade, weak gravitational lensing has flourished into a mature discipline for precision cosmology. Dedicated programs like the Dark Energy Survey (DES; \cite{DES2016}), the Kilo-Degree Survey (KiDS; \cite{DEJONG13}) or the Hyper Suprime-Cam Subaru Strategic Program (HSC-SSP; \cite{AIHARA18}) have measured the shapes of millions of galaxies in the sky to correlate their ellipticities  among themselves (cosmic shear) or with the position of foreground galaxies (galaxy-galaxy lensing shear). The joint analysis of such correlations along with galaxy clustering constitutes one of the most popular and powerful cosmological probes of modern astrophysics \citep{DESY3,AMON22,DESKIDS,ASGARI21,DVORNIK23,LI23}. Moreover, future measurements within the wide-area surveys conducted by the Large Synoptic Survey Telescope (LSST; \citep{LSST09}) and Euclid \citep{LAUREIJS11} will provide unprecedented precision for galaxy shape measurements and multi-band photometry.

Although it stands shoulder-to-shoulder with shear as equally fundamental gravitational lensing effects, magnification has been largely overlooked as an observable in itself and has been mostly studied as a correction to the true shear signals. The main reason for this has been attributed to the lower signal-to-noise ratio, which strongly depends on the (typically small) slope of the background population number counts. Indeed, cosmic magnification is the statistical manifestation of the gravitational lensing phenomenon of magnification bias \citep{SCHNEIDER92,BARTELMANN01}, consisting of two competing effects: around foreground overdensities, background sources are both magnified and diluted in the sky, creating a mismatch between the intrinsic and observed number counts of background sources whose magnitude is controlled by their number count slope. Whether this value is larger or smaller than unity dictates whether flux boosting dominates over solid angle dilution and determines both the sign and statistical significance of the angular correlation between the position of foreground galaxies (tracing matter overdensities) and background sources above a given flux limit. 

Despite observational hints as far back as the 1970s \citep{SELDNER79}, the first robust detection of cosmic magnification was published by \cite{SCRANTON05} using foreground galaxies and quasi-stellar objects (QSOs) from the Sloan Digital Sky Survey (SDSS; \cite{YORK00}). Since the slope of QSO number counts transitions from steep to flat as they become increasingly fainter, they found an angular correlation that shifted from positive to negative depending on the magnitude of the QSOs. However, only few papers followed along this line either using background QSOs \cite{MENARD10} or Lyman-break galaxies \citep{HILDEBRANDT09,MORRISON12}, none of which pursued a cosmological goal. Weak lensing  magnification was tacitly deemed inferior as a cosmological probe to tangential or cosmic shear measurements and it was not further pursued in large collaborations such as DES or KiDS, which concentrated on higher-S/N shear-based observables. 

However, the advent of the \emph{Herschel} space observatory \citep{PILBRATT10} marked a turning point for cosmic magnification. Equipped with the Spectral and Photometric Imaging REceiver (SPIRE; \cite{GRIFFIN10}) and the Photodetector Array Camera and Spectrometer (PACS; \cite{POGLITSCH10}), \emph{Herschel} provided photometric data in six bands (70, 100, 160, 250, 350 and 500 $\mu$m) and conducted two large-area surveys in the submillimeter, namely the \emph{Herschel} Multi-tiered Extragalactic Survey (HerMES; \cite{OLIVER12}) and the \emph{Herschel} Astrophysical Terahertz Large Area Survey (H-ATLAS; \cite{EALES10}), the latter covering the largest area, $\sim$ 660 deg$^2$. Although its original goal was to measure the dust content and the dust-obscured star formation for tens of thousands of local galaxies, a significant fraction of the sources detected by H-ATLAS was found to lie at high $(z>1)$ redshifts \citep{AMBLARD10,LAPI11,GON12,PEARSON13}. These were the high-redshift submillimeter galaxies (SMGs) first hinted at by the Submillimeter Common-User Bolometer Array (SCUBA) maps that had massive star formation rates and were nearly invisible in the optical band \citep{SMAIL97,BARGER98,HUGHES98,BLAIN02}. However, the H-ATLAS survey now provided wide-area catalogs of these sources for statistical studies \citep{VALIANTE16,SMITH03} and, in particular, for cosmic magnification.

Indeed, as first observed by SCUBA at 850 $\mu$m \citep{COPPIN06} or by the Balloon-borne Large Aperture Submillimeter Telescope (BLAST; \cite{DEVLIN09}) at 250, 350 and 500 $\mu$m \citep{PATANCHON09}, high-redshift SMGs feature
very steep number counts, a crucial factor for a high S/N cosmic magnification signal when used as a background sample. Moreover, due to the dust reprocessing of their ultraviolet and optical light, they are nearly transparent in these bands. As a consequence, cross-contamination between high-redshift SMGs and low-redshift lenses is not expected to be significant and both samples can, in principle, be cleanly separated. The first preliminary attempt to measure cosmic magnification on high-redshift SMGs detected by \emph{Herschel} was performed by \citep{WANG11} using early data $(\sim 13$ deg$^2$) from the HerMES survey. This was further refined by \cite{GON14} with much better statistics using the three equatorial fields surveyed by H-ATLAS $(\sim 160$ deg$^2$) and low-redshift galaxies from the Galaxy And Mass Assembly (GAMA) survey \citep{DRIVER09,BALDRY10,LISKE15}. In particular, \cite{GON14} found a highly significant GAMA/H-ATLAS angular correlation that could be accounted for by weak galaxy-galaxy lensing magnification. A follow-up work by \cite{GON17} extended the background sample to include additional H-ATLAS sources from a region in the Sotuh Galactic Pole, constituting a substantial improvement with respect to \cite{GON14} in terms of statistical significance. Moreover, the authors performed a tomographic analysis of the effect and interpreted the signal at small and intermediate scales via a halo model, which allowed them to characterize the typical halo masses of the intervening lenses. 

The aforementioned studies provided the foundation for cosmological work on cosmic magnification, a field that has not received much attention despite its latest developments. While early reservations on its (low) statistical significance were understandable twenty years ago, neglecting it with the current amount of astronomical data amounts to discarding valuable and additional physical information from an independent and complementary cosmological probe. What's more, high-redshift SMGs
constitute an optimal background sample for cosmic magnification studies and is yet to be fully exploited from the point of view of both simulations and observations.

This review summarizes the basic framework and state of the art in cosmic magnification studies using high-redshift SMGs for cosmology, both from the theoretical and observational points of view. Section \ref{sec2} outlines the theoretical framework underlying the observable, from the fundamental concept of magnification bias to its statistical quantification via the angular cross-correlation function. In Section \ref{sec3}, the standard methodology is described, including the typical source catalogs, the measurement approaches and the parameter estimation procedure. Section \ref{sec4} summarizes the latest results on cosmological constraints, both in a wide-bin and a tomographic setting, as well as parallel related studies. The outlook and challenges for cosmic magnification on high-redshift SMGs are discussed in Section \ref{sec5} and the conclusions are summarized in Section \ref{sec6}.

%%%%%%%%%%%%%%%%%%%%%%%%%%%%%%%%%%%%%%%%%%
\section{Theoretical basis}

In this section, the conceptual framework of cosmic magnification is outlined. The idea of magnification bias within the gravitational lensing context is presented, which is then extended in a statistical setting to cosmic magnification. The angular cross-correlation function is introduced as the observable quantifying this effect and a theoretical description of the galaxy-matter cross-correlation is given in terms of the halo model of structure formation.

\label{sec2}
\subsection{Magnification bias}

Gravitational lensing distorts and magnifies the light from background sources while conserving surface brightness. As a consequence, it has an impact on the perceived flux and size of the image(s). Let us denote by $n_{\text{b}_0}(S,z_s)$ and $n_{\text{b}_0}(>S_{\text{lim}},z_s)$ the intrinsic (unlensed) differential and integral number counts of background sources, respectively, that is,

\begin{equation}
    n_{\text{b}_0}(S,z_s)\equiv \frac{dN_{\text{b}_0}}{dS\,d\Omega \,dz_s}\quad\quad\quad  n_{\text{b}_0}(>S_{\text{lim}},z_s)=\int_{S_{\text{lim}}}^{\infty}dS \,n_{\text{b}_0}(S,z_s).
\end{equation}

\begin{figure}[H]
    \centering
    \includegraphics[width=\linewidth]{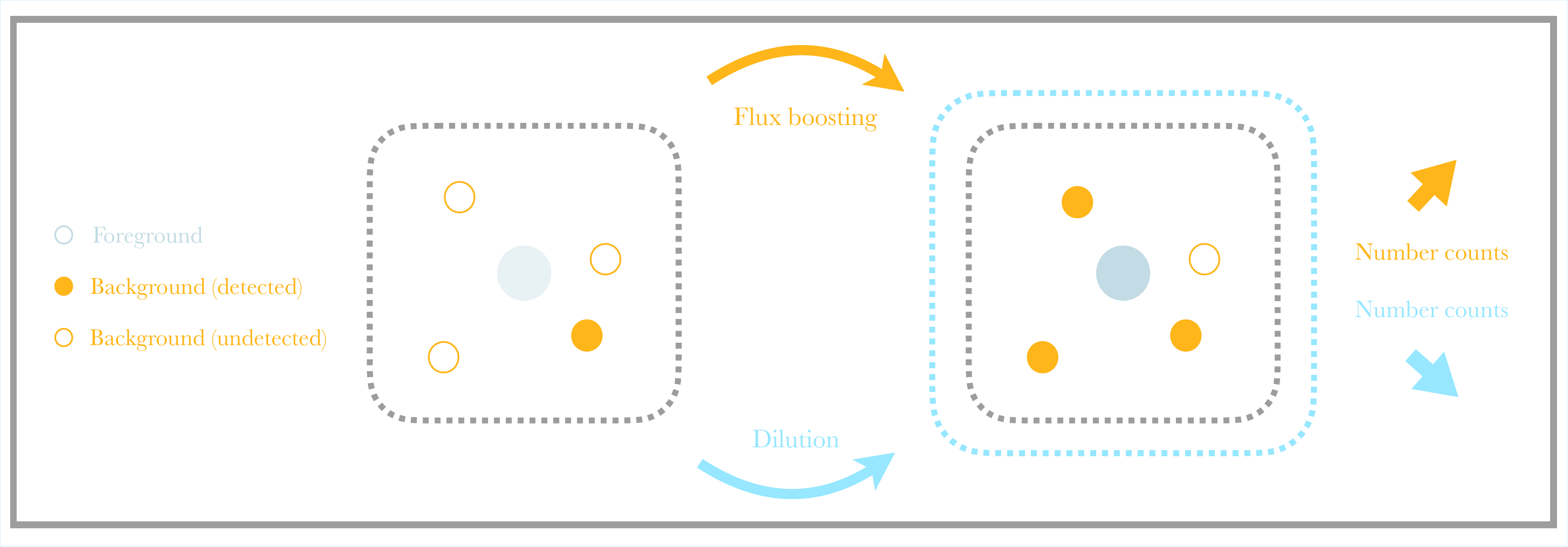}
    \caption{Schematic representation of magnification bias. While the flux boosting effect raises the observed number counts of background sources, the solid angle dilution reduces them, creating a mismatch with respect to the intrinsic number counts.}
    \label{magbias}
\end{figure}

Given a matter overdensity at redshift $z_l$ acting as a lens, the integral number counts of background sources around it are modified due to gravitational lensing following \citep{BARTELMANN01}

\begin{equation}
    n_{\text{b}}(>S_{\text{lim}},\vec{\theta})=\frac{1}{\mu(\vec{\theta})}n_{\text{b}_0}\bigg(>\frac{S_{\lim}}{\mu(\vec{\theta})}\bigg),
\end{equation}

where $\mu(\vec{\theta})$ is the magnification field at angular position $\vec{\theta}$ around the lens and the (lens and source) redshift dependence has been omitted for clarity. The above equation exemplifies the two effects that contribute to the modification of the observed number counts. On the one hand, the flux of background sources is boosted, which in turn raises the number counts since the threshold of detection is effectively lowered; in other words, sources that were intrinsically faint can now be detected. On the other hand, the solid angle subtended by background sources is stretched, essentially diluting the same sources over a larger area. The net effect modifies the intrinsic background number counts around matter overdensities and is termed magnification bias. A schematic picture of this effect is shown in Figure \ref{magbias}.

Whether flux boosting or dilution dominates over the other is strongly tied to the logarithmic slope of the number counts around the detection threshold. Indeed, assuming a power law behavior $n_{\text{b}_0}(>S)\propto S^{-\beta}$ in a neighborhood of $S_{\lim}$, it is straightforward to see that

\begin{equation}
    n_{\text{b}}(>S_{\text{lim}},\vec{\theta})=\mu^{\beta-1}(\vec{\theta})\,n_{\text{b}_0}(>S_{\lim}),
\end{equation}

which shows that $\beta>1$ ($<1$) produces an excess (defect) of background sources around foreground overdensities with respect to the absence of lensing.

\subsection{Cosmic magnification}

While the above equations are valid for a single lens, magnification bias can be studied in a cosmological setting by considering large samples of foreground tracers, which are then angularly correlated with background sources. In this scenario, magnification bias probes statistically the correlation between (lens) galaxies and the underlying matter field. In the weak lensing regime, this can be easily expressed in terms of the galaxy-matter cross-power spectrum, since

\begin{equation}
    \delta n_{\text{b}}(\vec{\theta};S_{\text{lim}})\equiv\frac{n_{\text{b}}(>S_{\text{lim}},\vec{\theta})-n_{\text{b}_0}(>S_{\text{lim}})}{n_{\text{b}_0}(>S_{\text{lim}})}=\mu^{\beta-1}(\vec{\theta})-1\approx 2(\beta-1)\kappa(\vec{\theta}),
\end{equation}

where $\kappa(\vec{\theta}))$ is interpreted as the effective convergence due to the large-scale structure in front of background sources. In this context, magnification bias manifests itself as a non-zero angular cross-correlation between two samples of galaxies at distinct redshifts:

\begin{equation}
    w_{\text{fb}}(\theta;S_{\text{lim}})\equiv\langle\delta n_{\text{f}}(\vec{\varphi})\delta n_{\text{b}}(\vec{\varphi}+\vec{\theta})\rangle_{\vec{\varphi}}=2(\beta-1)\langle \delta n_{\text{f}}(\vec{\varphi})\kappa(\vec{\varphi}+\vec{\theta})\rangle_{\vec{\varphi}}
\end{equation}

This cosmic magnification signal can be written under the Limber \citep{LIMBER53} and flat-sky approximations \citep{KAISER92} as \citep{COORAY02}

\begin{equation}
        w_{\text{fb}}(\theta;S_{\text{lim}})=2(\beta-1)\int_0^{\infty}\frac{dz}{\chi^2(z)}\frac{dN_{\text{f}}}{dz}W^{\text{lens}}(z)\int_0^{\infty}\frac{dl}{2\pi} l\,P_{\text{g-m}}(l/\chi(z),z)J_0(l\theta),
\end{equation}

where $\chi(z)$ is the comoving distance, $dN_{\text{f}}/dz$ is the normalized foreground redshift distribution, $P_{\text{g-m}}$ is the galaxy-matter power spectrum, $J_0$ is the zeroth-order Bessel function of the first kind and

\begin{equation}
    W^{\text{lens}}(z)\equiv \frac{3}{2}\frac{1}{c^2}\bigg[\frac{H(z)}{1+z}\bigg]^2\int_z^{\infty}dz'\frac{\chi(z)\chi(z'-z)}{\chi(z')}\frac{dN_{\text{b}}}{dz'}
\end{equation}

is the so-called lensing kernel.

\subsection{Description of the galaxy-matter cross-correlation: the halo model}

The correlation between galaxies and the underlying matter density field in the cosmic magnification signal is quantified via the galaxy-matter cross-power spectrum. For large scales, one can use a linear galaxy bias model, for which
\begin{equation}
    P_{\text{g-m}}(k,z)\approx b_1(z)P_{\text{mm}}(k,z),
\end{equation}
and then compute the matter power spectrum with non-linear corrections using, for instance, empirical fitting functions like HALOFIT \citep{SMITH03,TAKAHASHI12}. Approaches based on higher-order perturbation theory enable extending the validity range to mildly non-linear scales \citep{MCDONALD09,SAITO14}. However, cosmic magnification studies have relied on a description of the galaxy matter correlation based on the halo model of structure formation \citep{SCHEBERT91,SEL00,COORAY02,ASGARI23}. 

The halo model constitutes an analytical framework to derive the statistical properties of matter and its tracers provided that its ingredients are correctly calibrated, typically via N-body dark matter simulations. In its base form, the underlying assumption is that all mass in the Universe is bound up into distinct linearly biased halos whose properties are only a function of their mass. Within this approach, the prescription for the galaxy-matter correlation can be split into a one-halo and a two-halo term.

\begin{equation}
    P_{\text{g-m}}(k,z)=P_{\text{g-m}}^{\text{1h}}(k,z)+P_{\text{g-m}}^{\text{2h}}(k,z),
\end{equation}
where

\begin{align}
    P_{\text{g-m}}^{\text{1h}}(k,z)&=\int_0^{\infty} dM_h\,M_h\frac{n(M_h,z)}{\bar{\rho}_{\text{m}}(0)}\frac{1}{\bar{n}_g(z)}\bigg[\langle N_c\rangle_{M_h}\,|u(k|M_h,z)|+\langle N_s\rangle_{M_h}\,\,|u(k|M_h,z)|^2\bigg]\\
    P_{\text{g-m}}^{\text{2h}}(k,z)&=P^{\text{lin}}(k,z)\Bigg[\int_0^{\infty}dM_h\,M_h\frac{n(M_h,z)}{\bar{\rho}_{\text{m}}(0)}b^{\text{h}}_1(M_h,z)u(k|M_h,z )\Bigg]\,\cdot\nonumber\\
    &\cdot\Bigg[\int_0^{\infty}dM_h\frac{n(M_h,z)}{\bar{n}_g(z)}b_1(M_h,z)\,\Big(\langle N_c\rangle_{M_h} + \langle N_s \rangle_{M_h} \,u(k|M_h,z)\Big)\Bigg].
\end{align}

In the above equations, $n(M_h,z)$ is the halo mass function, $b_1(M_h,z)$ is the linear deterministic halo bias, $\bar{n}_g(z)$ is the mean number density of galaxies, $u(k|M_h,z)$ is the normalized Fourier transform of the dark matter density profile, $\bar{\rho}_\text{m}(0)$ is the background matter density at $z=0$ and $\langle N_i\rangle_{M_h}$  is the mean number of central $(i=c)$ and satellite $(i=s)$ galaxies in a halo of mass $M_h$, that is, the Halo Occupation Distribution. 

The halo mass function is usually parametrized via

\begin{equation}
    n(M_h,z)=\frac{\bar{\rho}_{\text{m}}(0)}{M_h^2}f(\nu)\bigg|\frac{d\log \nu}{d\log M_h}\bigg|,
\end{equation}

where

\begin{equation}
    \nu(M_h,z)\equiv \bigg[\frac{\hat\delta_c(z)}{\sigma(M_h,z)}\bigg]^{2}
\end{equation}

is the so-called peak height, $\hat{\delta}_c(z)$ is the linear overdensity threshold at redshift $z$ for a halo to collapse at that same redshift \citep{KITAYAMA96} and $\sigma^2(M_h,z)$ is the variance of the smoothed linear overdensity field over a scale $R$ corresponding to the mass $M_h=4\pi/3\,\bar{\rho}_{\text{m}}(0)R^3$. The function $f(\nu)$ depends on the adopted halo mass function model. For instance, the Sheth-Tormen model \citep{ST99} reads

\begin{equation}
    f_{\text{ST}}(\nu;A,a,p)=A\sqrt{\frac{a\nu}{2\pi}}\bigg[1+\bigg(\frac{1}{a\nu}\bigg)^p\bigg]e^{-a\nu/2},
\end{equation}
where its parameters are typically derived from a best fit to dark matter N-body simulations: $A=0.322$, $a=0.707$ and $p=0.3$.

The deterministic linear halo bias can be computed from the halo mass function via the peak-background split argument \citep{COLE89,MO96}, where the density field is divided into a large-scale and a small-scale component. In particular, for a Sheth-Tormen halo mass function model, the linear halo bias reads
\begin{equation}
    b^{\text{h}}_{1_{\text{ST}}}(M_h,z)=1+\frac{1}{\hat{\delta}_c(z)}\bigg(a\nu-1+\frac{2p}{1+(a\nu)^p}\bigg).
\end{equation}

Regarding the dark matter density profile, a Navarro-Frenk-White (NFW; \cite{NAVARRO97}) model is typically adopted, for which its normalized Fourier transform can be analytically computed as

\begin{align}
    u_{\text{NFW}}(k|M_h,z)&=\frac{1}{\log(1+c)-c/(1+c)}\Big[\sin kr_s[\text{Si}([1+c]kr_s)-\text{Si}(kr_s)]-\frac{\sin ckr_s}{[1+c]kr_s}+\nonumber\\
    &+\cos kr_s[\text{Ci}([1+c]kr_s)-\text{Ci}(kr_s)\Big],
\end{align}

where $c(M_h,z)$ is a halo mass-concentration relation \citep{BULLOCK01,DUFFY08,DUTTON14,DIEMER15} and $r_s=R_h/c$ is the scale radius, where $R_h$ is the halo radius, defined via a given overdensity criterion. For a virial overdensity spherical-collapse computation \citep{WEINBERG03} 

\begin{equation}
    M_h=\frac{4}{3}\pi \Delta_{\text{vir}}(z)\bar{\rho}_{\text{m}}(z) R_h^3,
\end{equation}

where 

\begin{equation}
    \Delta_{\text{vir}}(z)=18\pi^2\bigg(1+\frac{0.399}{[1/\Omega_m(z)-1]^{0.941}}\bigg).
\end{equation}

Lastly, regarding the galaxy-halo connection, the halo model relies on halo occupation distribution (HOD) models, which describe the average number of galaxies in a halo in terms of its mass, typically split into the contribution of central and satellite galaxies. Studies on cosmic magnification with background SMGs have so far employed the simple three-parameter $(\alpha, M_{\text{min}},M_1)$ model of \cite{ZEHAVI05}, which reads

\begin{equation}
    \langle N\rangle_{M_h}=\langle N_c\rangle_{M_h}+\langle N_s \rangle_{M_h}=\bigg[1+\bigg(\frac{M_h}{M_1}\bigg)^{\alpha}\bigg]\Theta(M_h-M_{\text{min}}),
\end{equation}

where $M_{\text{min}}$ is the average minimum mass for a halo to host a (central) galaxy, $M_1$ is the average halo mass at which there is exactly one satellite and $\alpha$ is the logarithmic slope of the satellite occupation number. Other options include the five-parameter model of \cite{ZHENG05} or an approach based on the conditional luminosity function \citep{YANG08}.

Despite its versatility and success, the halo model has a number of limitations that cannot go unnoticed \citep{ASGARI23}. In particular, its "vanilla" version differs up to 15\% from semi-analytical theoretical calculations from the DARKEMU simulation-based emulator \citep{NISHIMICHI19}, especially in the transition between the one- and two-halo regimes. The inclusion of non-linear halo bias from simulations \citep{MEAD21}, which in turn incorporates halo exclusion, slightly alleviates these differences. However, the most accurate version of the halo model is that of HMCODE \citep{MEAD15,MEAD20,MEAD21B}, which adds physically motivated free parameters to the base model that are fit to numerical simulations to capture non-linear and baryonic effects. Lastly, simulation-based emulators \citep{HEITMANN09,HEITMANN10,LAWRENCE10,HEITMANN16,LAWRENCE17,KNA19,KNA21,ANGULO21,CONTRERAS24} are becoming popular in the literature and should be considered in future work on cosmic magnification, since they allow for fast and accurate computations of non-linear correlations of cosmological fields, although their range of validity for different cosmologies must be taken with care.

%%%%%%%%%%%%%%%%%%%%%%%%%%%%%%%%%%%%%%%%%%
\section{Methodology}
\label{sec3}

This section is devoted to the description of the standard methodology in cosmic magnification studies with background SMGs. This involves the choice of galaxy samples, the procedure for the measurement of the angular cross-correlation function and the estimation of the model parameters, including those relevant for cosmology.

\subsection{Galaxy samples}

The submillimeter sky has been largely overlooked in terms of wide-area surveys, with the only exceptions of the HerMES and H-ATLAS programs carried out by the \emph{Herschel} space observatory. Due to its larger area, cosmic magnification studies on background SMGs have mostly employed sources detected by H-ATLAS, which surveyed several fields chosen to minimize dust emission in the Milky Way and to maximize the complementary data from other wavelengths. This resulted in five regions: one close to the North Galactic Pole (NGP), three on the celestial equator (G09, G12 and G15) and one near the South Galactic Pole (SGP), amounting to a total of $\sim$ 660 deg$^2$, although only the parts overlapping with foreground samples have been essentially used. Using the peak of dust emission, photometric redshifts were computed via a $\chi^2$ fit to the spectral energy distribution of SMM J2135-0102, a gravitationally-lensed SMG at $z=1.2$ that was found to provide the best overall fit to H-ATLAS data \citep{LAPI11,GON12,IVISON16}. The background sample is then typically restricted to $1.2<z<4.0$ to avoid any possible contamination with foreground objects. To take into account errors in photometric redshift estimation, the fiducial redshift distribution derived from the SED fitting, $\mathcal{P}^{\text{fid}}_{\text{b}}$, is weighted as
\begin{equation}
    \mathcal{P}_{\text{b}}(z)=\mathcal{P}^{\text{fid}}_{\text{b}}(z)\int d z_{\text{ph}}\,W(z_{\text{ph}}) \,\mathcal{P}(z_{\text{ph}}|z),
\end{equation}

where 

\begin{align}
    W(z_{\text{ph}})=\begin{cases}
    1 \quad \text{if} \quad z_{\text{ph}}\in[1.2,4.0]\\
    0 \quad \text{if} \quad z_{\text{ph}}\notin[1.2,4.0]
    \end{cases}
\end{align}
is the window function and $\mathcal{P}(z_{\text{ph}}|z)$ is the photometric redshift error function, parametrized as a Gaussian distribution with zero mean and dispersion $(1+z)\sigma_{\Delta z/(1+z)}$, for which the value $\sigma_{\Delta z/(1+z)}=0.153$ was adopted, as found by \cite{IVISON16}.

Regarding the lenses, and due to the coordination with H-ATLAS, most works on cosmic magnification with SMGs have relied on the GAMA II survey \citep{DRIVER11,BALDRY10,LISKE15}, a spectroscopic follow-up of the imaging of SDSS, KiDS and other surveys down to a limiting magnitude of $r<19.8$. In particular, the regions overlapping with the H-ATLAS survey amount to a total of $\sim$ 207 deg$^2$ and contain $\sim$ 130 000 galaxies with spectroscopic redshifts in the range $0.2<z<0.8$. Depending on the scope of the work, this sample has also been further subdivided into clean redshift bins for a tomographic analysis. Additional foreground samples employed by related studies include galaxy clusters detected from the SDSS-III \citep{WEN12} or DESI legacy imaging \citep{ZOU21} surveys or the QSO catalog from the 12th data release of SDSS \citep{paris17}. In all cases, redshifts cuts were applied so as to ensure no overlap with background SMGs.

\subsection{Measurements}

One of the advantages of cosmic magnification lies in the simplicity of its measurement. Unlike shear-based observables, sources need not be resolved, no point spread function corrections are required and there is no need for intrinsic alignment corrections or shear calibration. Cosmic magnification directly deals with the number of foreground-background pairs at a given angular distance, for which accurate photometry and source detection are the only requisites. Since it is quantified via the excess or defect of background sources around foreground galaxies with respect to the absence of lensing, it can be quantified via a Landy-Szalay-like \citep{LANDY93,HERRANZ01} estimator, that is,

\begin{equation}
    \hat{w}_{\text{fb}}(\theta)=\frac{\text{D}_{\text{f}} \text{D}_{\text{b}}(\theta)-\text{D}_{\text{f}}\text{R}_{\text{b}}(\theta)-\text{D}_{\text{b}}
    \text{R}_{\text{f}}(\theta)+\text{R}_{\text{f}}\text{R}_{\text{b}}(\theta)}{\text{R}_{\text{f}}\text{R}_{\text{b}}(\theta)}
\end{equation}

where $\text{X}_{\text{f}} \text{Y}_{\text{b}}(\theta)$ is the normalized number of foreground-background pairs at an angular distance $\theta$; when X$\equiv$D, the galaxies are chosen from the data, whereas X$\equiv$R implies that the sources are selected from a randomly (unclustered) generated catalog. It should be noted that the random catalogs need to reflect the selection function of the samples in order to yield an unbiased measurement. Indeed, cosmic magnification requires control over the angular completeness of the catalogs reflecting spatial fluctuations in depth or detection efficiency. For example, in the case of the GAMA survey, random catalogs are available for two-point correlation function analyses, while for H-ATLAS instrumental noise maps can be imprinted on the random catalog to account for the scanning strategy.

Although the estimator has remained unchanged across all studies, the measurement approach differs between older and more recent papers on cosmic magnification with SMGs. In particular, since the GAMA/H-ATLAS data consisted of four spatially-separated regions, the strategy initially involved a "tile-and-average" strategy, averaging over cross-correlation measurements performed in distinct patches, whether these were the full fields or equal-area subdividisions (tiles). The approach was later refined to provide a single measurement of the angular cross-correlation by adding up all pairs from the different regions. Since different scales are correlated, this requires a quantification of the covariance matrix for the measurement. While this is straightforward via the sample covariance for the former measurement strategy, a Bootstrap approach is adopted for the most recent methodology. This involves dividing the whole area into $N$ subregions, resampling $N_r$ of them with replacement and repeating the process $N_b$ times. A measurement of the cross-correlation is performed on each of the $N_b$ Bootstrap samples and the covariance matrix is given by

\begin{equation}
    \text{Cov}(\theta_i,\theta_j)=\frac{1}{N_b-1}\sum_{k=1}^{N_b}\,\bigg[\hat{w}_k(\theta_i)-\bar{\hat{w}}(\theta_i)\bigg]\bigg[\hat{w}_k(\theta_j)-\bar{\hat{w}}(\theta_j)\bigg]\label{covariance},
\end{equation}

where $\hat{w}_k$ denotes the measured correlation function from the $k^{\text{th}}$ Bootstrap sample and $\bar{\hat{w}}$ is the corresponding average value over all Bootstrap samples.

\begin{figure}[H]
    \centering
    \includegraphics[width=0.8\linewidth]{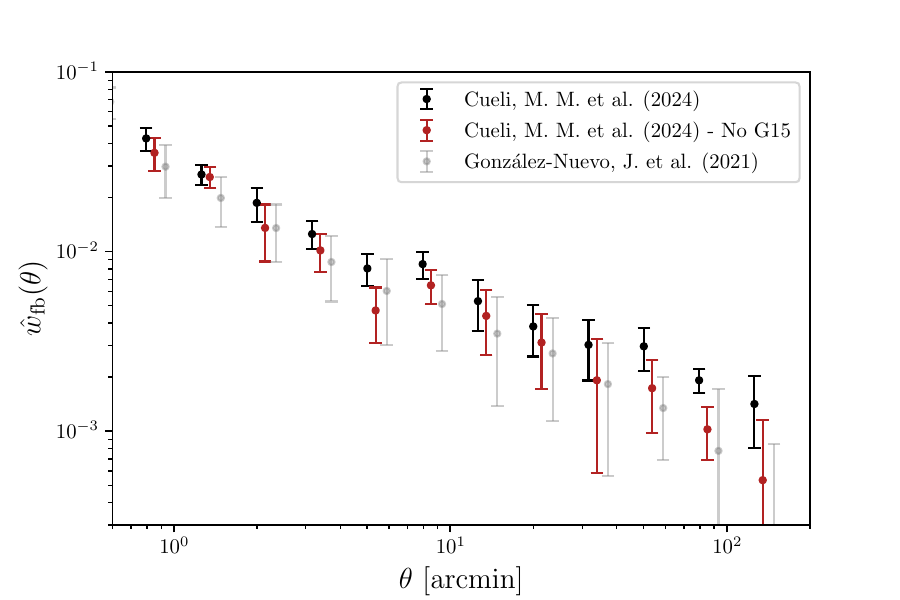}
    \caption{GAMA/H-ATLAS cross-correlation data. The grey data points show the measurements from the "tile-and-average" approach used in \cite{GON21}, while the black and red data depict the single-measurement strategy using all fields and exlucding the G15 region, respectively \citep{CUE24a}. This figure has been modified with respect to the original source by \cite{CUE24a}.}
    \label{xc_data}
\end{figure}

With regard to the choice of $N_r$, that is, the number of subregions that are drawn (with replacement) for each Boostrap sample, $N_r=3N$ has been adopted following the conclusions of \cite{NORBERG09}, since they obtained an excellent agreement with external estimates of the errors in this case. For the GAMA/H-ATLAS analyses, each independent field is divided into five patches, thus fixing $N=20$. The procedure was then repeated $N_b=10000$ times for stability. Figure \ref{xc_data} depicts the GAMA/H-ATLAS data from \cite{CUE24a}, that used this approach, compared to the "tile-and-average" strategy of \cite{GON21}, shown in red.

\subsection{Parameter estimation}

The angular cross-correlation measurements are then fit to the cosmic magnification prediction, typically within the halo model formalism. Within this approach, the posterior distribution of the relevant parameters entering the theoretical description of the signal is sampled via a Markov chain Monte Carlo approach. Denoting by $\{p_j\}_{j=1}^{n}$ the set of parameters to be fit, the log-likelihood can be described by an $n$-dimensional multivariate Gaussian, that is,

\begin{align}
    \log{\mathcal{L}\,(\theta_1,\ldots,\theta_m;\{p_j\}_{j=1}^n)}=-\frac{1}{2}&\bigg[m\log{(2\pi)}+\log{|C|}+\overrightarrow{\varepsilon}^{\text{T}}C^{-1}\,\overrightarrow{\varepsilon}\bigg],
\end{align}

where 

\begin{equation}
    \overrightarrow{\epsilon}\equiv [w_{\text{fb}}(\theta_1;\{p_j\}_{j=1}^n))-\hat{w}_{\text{fb}}(\theta_1),\ldots,w_{\text{fb}}(\theta_m;\{p_j\}_{j=1}^n))-\hat{w}_{\text{fb}}(\theta_m)],    
\end{equation}

and $C$ is the covariance matrix associated to the cross-correlation measurements.

Within the halo model interpretation of the cosmic magnification signal, the relevant parameters to be fit are the ones describing cosmology, the HOD and the logarithmic slope of the SMG number counts. Regarding cosmology, just like shear-based observables, cosmic magnification is mostly sensitive to the matter density parameter ($\Omega_m$) and the amplitude of the smoothed matter overdensity field ($\sigma_8$), although the Hubble constant was also allowed as a free parameter via $H_0=100h$ km$\,·s^{-1}·$Mpc$^{-1}$. For a $\Lambda$CDM model, the rest of parameters are fit to \emph{Planck}'s best-fit values \citep{PLANCKVIII20}, that is, $\Omega_b=0.0486$ and $n_s=0.967$. However, when testing dynamical dark energy models under the Chevallier-Polarski-Linder parametrization of the equation of state \citep{CHEVALL01,LINDER03}, the $w_0$ and $w_a$ parameters are also fit to the data. While the prior distributions for the cosmological parameters have not remained unchanged across cosmic magnification papers, uniform distributions were chosen for all of them featuring a wide enough range, typically:

\begin{align}
    \Omega_m\sim\mathcal{U}\,[0.1,1.0]\,\,&\quad\quad \sigma_8\sim\mathcal{U}\,[0.6,1.2]\quad\quad h\sim\mathcal{U}\,[0.5,1.0]
\end{align}
\begin{align}
    w_0\sim\mathcal{U}\,[-2.0,0.0]\quad\quad w_a\sim\mathcal{U}\,[-3.0,3.0].
\end{align}
    
Concerning the typically adopted three-parameter HOD model, the prior distributions for its parameters are again uniform with reasonably large ranges, namely

\begin{align}
    \alpha\sim\mathcal{U}\,[0.1,1.5]\quad\quad \log \bigg[\frac{M_{\text{min}}}{h^{-1}M_{\odot}}\bigg]\sim\mathcal{U}\,[10,16]\quad\quad \log \bigg[\frac{M_{\text{1}}}{h^{-1}M_{\odot}}\bigg]\sim\mathcal{U}\,[10,16]
\end{align}

Lastly, the logarithmic slope of the SMG number counts was fixed to $\beta=3$ in the first papers on cosmic magnification on account of the results found by \cite{LAPI11}. In that work, the observed number counts of high-redshift \emph{} SMGs were correctly reproduced via an updated version of the galaxy formation model by \cite{LAPI06}, which predicted a logarithmic slope of $\beta\sim3$ for massive proto-spheroids at 350 $\mu$m at the 3$\sigma$ detection limit. However, later work by \cite{CUE24a} performed a more thorough analysis that yielded a more reasonable prior for this parameter featuring a Gaussian distribution around the value 2.9, that is, 
\begin{equation}
    \beta\sim\mathcal{N}\,[2.90,0.04].
\end{equation}

In summary, in the cases considered up to now with cosmic magnification studies on SMGs, the set of parameters to be fit is given by

\begin{equation}
    \{p_j\}_{j=1}^{n}=\{\Omega_m,\sigma_8,h,w_0,w_a,\alpha,M_{\text{min}},M_1,\beta\}.
\end{equation}

It should be noted that a tomographic analysis in different lens redshift bins requires a further subdivision of the HOD parameters to describe the redshift evolution of the galaxy-halo connection.

\section{Results}
\label{sec4}

This section summarizes the main results involving cosmic magnification on high-redshift SMGs. These have been divided into wide-bin and tomographic studies according to the characterization of the lens sample and include papers focused on both methodology and cosmological constraints. A number of parallel works involving the halo mass function and the small-scale behavior of magnification will also be discussed. It should be noted that all these results involve the sole use of cosmic magnification, without the combination with other external probes.

\subsection{Cosmology using a single wide redshift bin}

Although it is physically informative to divide the lens sample for a tomographic analysis, a higher signal-to-noise ratio can be attained if all lenses are combined into a single redshift bin for an initial study. Along this line, \cite{BON20} provided a first a proof-of-concept paper to demonstrate the constraining power of cosmic magnification. Indeed, cosmic magnification (via the angular cross-correlation function) was shown to be sensitive to both $\Omega_m$ and $\sigma_8$, whereas the dependence on the Hubble constant was much weaker, as expected from a projected lensing observable. Within the $\Lambda$CDM framework, they derived a mean value of $\sigma_8=0.78^{+0.07}_{-0.15}$ and a lower limit of $\Omega_m>0.24$ at 95\% credibility. In addition, these parameters were shown not to display the degeneracy direction typically found in shear observables.

However, the preference for very high values of $\Omega_m$ found by \cite{BON20} was soon showed to be biased due to the measurement methodology. Even if the total area was $\sim$ 207 deg$^2$, the angular cross-correlation had been measured and averaged over a number of subregions of $\sim$ 4 deg$^2$. As shown by \cite{GON21}, an integral constraint correction was needed to compensate for the loss of galaxy pairs due to the (small) finite size of the  subregions. Both this and a correction for the large-scale fluctuations imprinted onto the data by the H-ATLAS scanning strategy were implemented in the measurements by \cite{GON21}, which reanalyzed the data. These changes translated into an enhancement of the cross-correlation signal at the highest angular distances, primarily absorbed by the matter density parameter, disfavoring extremely high values. In essence, they found mean values of $\Omega_m=0.45^{+0.13}_{-0.21}$ (with a mode of 0.38) and $\sigma_8=0.84^{+0.11}_{-0.18}$ and confirmed the behavior on the $\Omega_m-\sigma_8$ plane anticipated by \cite{BON20}.

Despite the effort, the cross-correlation data at the largest scales were shown to be sensitive to the exact value of the integral constraint correction, the computation of which is not completely unbiased. In order to circumvent this problem, the measurement methodology was refined in \cite{CUE24a} by considering a single measurement of the angular cross-correlation via the sum of all galaxy pairs across all regions, as detailed in Section \ref{sec3}, along with a covariance based on the Bootstrap method with a resampling factor. This approach yielded a dataset with smaller uncertainties and, since there was no longer any limitation due to the size of subregions, it enabled access to larger angular scales without the need for an integral constraint correction. In addition to this, \cite{CUE24a} underlined the importance of the prior information on the $\beta$ parameter (describing the logarithmic slope of the background integral number counts) for obtaining unbiased cosmological constraints.
By assuming a wide uniform prior on this parameter, they found that, whereas its value barely had an impact on the matter density parameter, it was heavily degenerate with $\sigma_8$. However, the agreement on the value of $\beta$ between the direct measurement from the number counts at the detection threshold and the prediction from the galaxy formation model of \cite{LAPI11} (which correctly reproduced the observed H-ATLAS data) led to the choice of a Gaussian prior around 2.9 for high-redshifts SMGs.

\begin{figure}[h]
    \centering
    \includegraphics[width=0.6\linewidth]{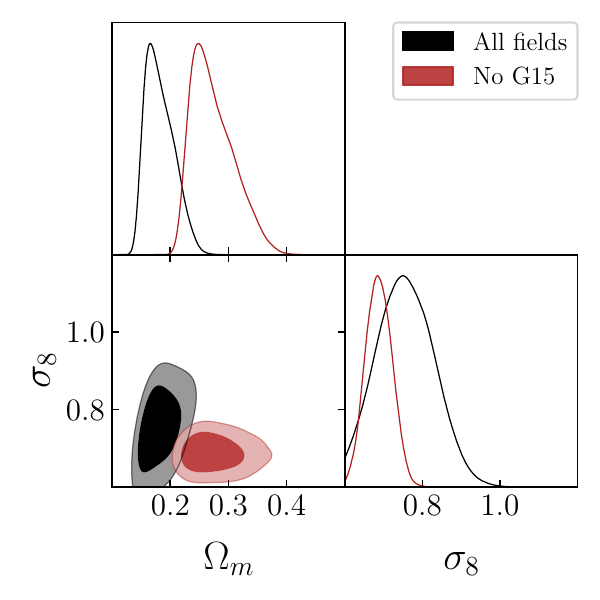}
    \caption{Posterior distributions of $\Omega_m$ and $\sigma_8$ from \cite{CUE24a} after marginalization over the rest of parameters. Black and red plots denote the results using all four GAMA/H-ATLAS regions and excluding G15, respectively. This figure has been modified with respect to the original source by \cite{CUE24a}.}
    \label{corner_1}
\end{figure}

Using this new dataset and the prior information on $\beta$, \cite{CUE24a} performed a cosmological analysis within the $\Lambda$CDM model, obtaining mean values of $\Omega_m=0.23^{+0.03}_{-0.06}$ and $\sigma_8=0.79^{+0.10}_{-0.10}$, as shown in black in Figure \ref{corner_1}. However, the relatively low value of $\Omega_m$ sparked further analysis on the large-scale behavior of the signal. Indeed, by measuring cosmic magnification in each of the four regions separately, the authors showed that G15 displayed an excess of cross-correlation with respect to the other zones, especially at the largest scales, where it remained practically flat. Given that the source selection criteria, redshift and (foreground) stellar mass distributions were remarkably uniform across all four regions, the authors concluded that this difference was physical and associated to sampling variance, especially taking into account that the selection function of both catalogs had been carefully included. 

The importance of this issue is emphasized quantitatively when excluding the G15 region from the analysis, since the mean values of the two above cosmological parameters are shifted to $\Omega_m=0.27^{+0.03}_{-0.04}$ and $\sigma_8=0.72^{+0.04}_{-0.04}$, as depicted in red in Figure \ref{corner_1}. This behavior highlights the importance of addressing physical differences among spatial (sub)regions that, in principle, should be homogeneously surveyed and large enough to derive unbiased cosmological constraints. In fact, \cite{CUE24a} also observed important discrepancies in the angular clustering of GAMA galaxies across all four regions of the sky. Similar differences were also noted by \cite{FER24}, where galaxy clusters were employed as lenses to study cosmic magnification on H-ATLAS SMGs. Capitalizing on the large masses of these systems, the authors obtained a robust measurement of magnification bias and characterized the HOD of galaxy clusters, obtaining typical minimum masses of $\sim 10^{13} M_{\odot}/h$ for cluster-sized halos. However, the largest scales of the signal were shown to reproduce an excess of cross-correlation mainly due to the G15 region, which in turn yielded typically low values for the matter density parameter, in agreement with the above discussions using galaxy samples as lenses.

Lastly, the potential of cosmic magnification to constrain the sum of neutrino masses was studied in \cite{CUE24B}, since the free-streaming scale set by the non-zero mass of these particles imprints a scale-dependent suppression in the galaxy-matter power spectrum. The authors found that the
sensitivity of the observable varied significantly depending on whether the cosmological model was parametrized in terms of the amplitude of the primordial power spectrum $(A_s)$ or of the root mean square of the $z=0$ smoothed overdensity field $(\sigma_8)$. In particular, the effect of varying neutrino masses was shown to be concentrated on small and intermediate scales when $A_s$ was kept fixed. Capitalizing on this, the GAMA/H-ATLAS cross-correlation data enabled an upper limit of $\sum m_{\nu}<0.78$ eV at 95\% credibility for an $A_s$ value fixed to the massless neutrino scenario, a situation that qualitatively remained unchanged with a Gaussian prior on this parameter. Similarly to previous works, the influence of the G15 region is visible, with a mean matter density parameter transitioning from $\Omega_m=0.18^{+0.02}_{-0.02}$ to $\Omega_m=0.34^{+0.02}_{-0.03}$ after its exclusion from the analysis. However, the slightly larger error bars when only three fields are present hinder the possibility to constrain neutrino masses, marking a minimum signal-to-noise ratio needed for this task.

%%%%%%%%%%%%%%%%%%%%%%%%%%%%%%%%%%%%%%%%%%%%%%%%%%
%%%%%%%%%%%%%%%%%%%%%%%%%%%%%%%%%%%%%%%%%%%%%%%%%%
\subsection{Cosmology with a tomographic analysis}
%%%%%%%%%%%%%%%%%%%%%%%%%%%%%%%%%%%%%%%%%%%%%%%%%%
%%%%%%%%%%%%%%%%%%%%%%%%%%%%%%%%%%%%%%%%%%%%%%%%%%

The redshift behavior of the cosmic magnification signal has also been studied in a tomographic setup by dividing the GAMA lens sample in a number of redshift bins. This approach aims to deproject the observable and enables a more accurate characterization of redshift-dependent phenomena like the potential time evolution of dark energy or the galaxy-halo connection. While the number of GAMA sources is much lower than alternative photometric catalogs, the spectroscopic nature of its redshifts enables a clean separation into non-overlapping subsamples, effectively removing any uncertainty on the lens redshift distribution.

Although tomography in this context was originally introduced in \cite{GON17}, the first cosmological constraints derived from  measurements of GAMA/H-ATLAS cross-correlations were published by \cite{BON21}, which incorporated the large-scale corrections outlined in \cite{GON21}. Under the "tile-and-average" approach, the authors measured cosmic magnification on high-redshift SMGs in four redshift bins between 0.1 and 0.8 and modeled the observable in $\Lambda$CDM, $w_0$CDM and $w_0w_a$CDM cosmologies with a different set of parameters for the lens HOD in each bin. In the first case,
the authors obtained mean values of $\Omega_m=0.33^{+0.08}_{-0.16}$ and $\sigma_8=0.87^{+0.13}_{-0.12}$, in agreement with the nontomographic results from \cite{GON21}, obtained under the same standard cosmological model. The inclusion of a potential dynamical dark energy within the CPL parametrization barely modified these constraints and additionally yielded mean values of $w_0=-1.00^{+0.53}_{-0.56}$ for the $w_0$CDM model and $w_0=-1.09^{+0.75}_{-0.63}$ and $w_a=-0.19^{+1.67}_{-1.69}$ for the $w_0w_a$CDM model. Despite the relatively large uncertainties, the central values showed a remarkable agreement with a cosmological constant ($w_0=-1$ and $w_a=0$).

\begin{figure}[H]
  \centering
  % first row, left
  \begin{minipage}[b]{0.51\linewidth}
    \includegraphics[width=\linewidth]{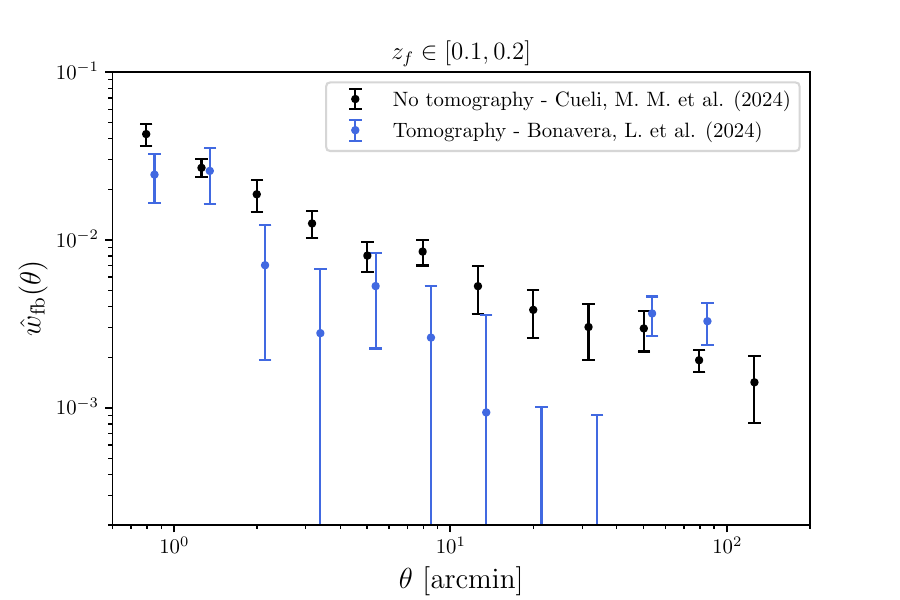}
    \label{fig:1a}
  \end{minipage}
  \hspace{-4ex}
  % first row, right
  \begin{minipage}[b]{0.51\linewidth}
    \includegraphics[width=\linewidth]{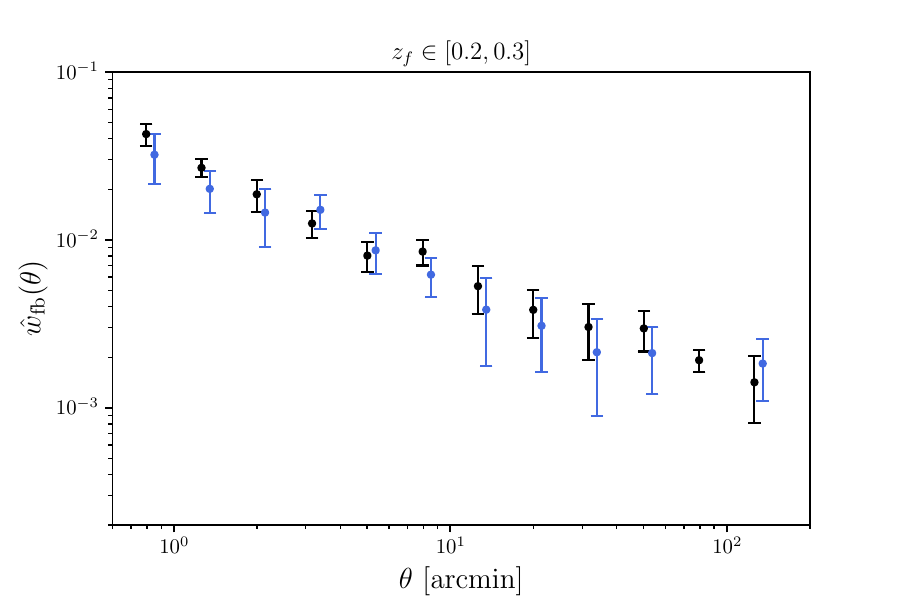}
    \label{fig:1b}
  \end{minipage}

  \vspace{-2.6ex}   % small vertical gap between rows

  % second row, left
  \begin{minipage}[b]{0.51\linewidth}
    \includegraphics[width=\linewidth]{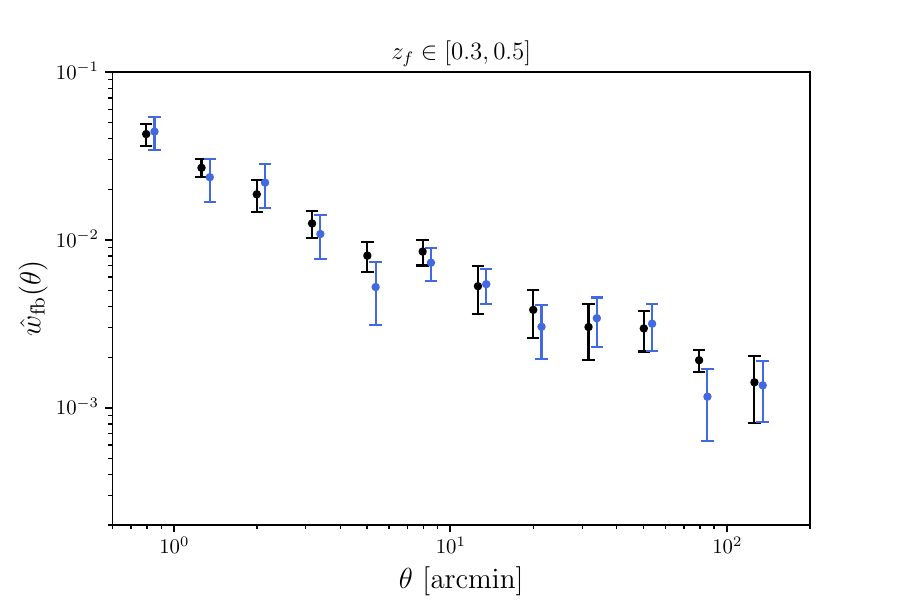}
    \label{fig:1c}
  \end{minipage}
  \hspace{-4ex}
  % second row, right
  \begin{minipage}[b]{0.51\linewidth}
    \includegraphics[width=\linewidth]{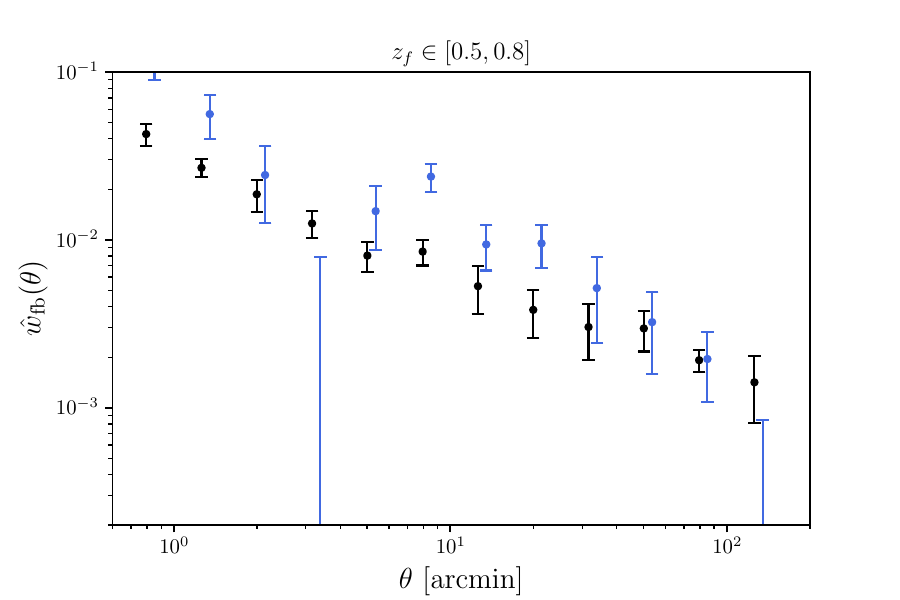}
    \label{fig:1d}
  \end{minipage}
  \vspace{-3pt}
  \caption{GAMA/H-ATLAS cross-correlation data in the tomographic setting. The blue data points depict the measurements in each of the four lens redshift bins chosen by \cite{BON24}, namely $z_f\in[0.1,0.2]$, $[0.2,0.3]$, $[0.3,0.5]$ and $[0.5,0.8]$. The single-bin data from \cite{CUE24a} are shown in black for comparison. This figure has been modified with respect to the original source by \cite{BON24}.}
  \label{xc_data_tomography}
\end{figure}

Analogously to the nontomographic setup, the analysis was refined by \cite{BON24} within $\Lambda$CDM using a single cross-correlation measurement across the whole area and a Bootstrap covariance estimation. The measurements are plotted in blue in Figure \ref{xc_data_tomography} for a visual comparison with the non-tomographic approach of \cite{CUE24a}, depicted in black. The results were qualitatively similar to the wide redshift bin analysis of \cite{CUE24a}: at every redshift bin, the data error bars are significantly reduced and an excess signal is found, especially at the largest scales, due to the presence of the G15 region. This behavior induced even lower values of the matter density parameter with respect to \cite{CUE24a}, with a mode of 0.17. However, this value increased to 0.32 when G15 was excluded from the analysis, extending the conclusions from the nontomographic study to an effect observed at different redshifts. In addition, the possibility of splitting the lens sample in six (instead of four) redshift bins was explored, as well as extending the range slightly further. The authors found that the latter choice seemed to reduce the effect of sampling variance and alleviate its impact on the matter density parameter.

\subsection{Parallel analyses}

Aside from exploiting magnification bias on background SMGs to constrain cosmology, several other goals of astrophysical interest can be pursued by fixing the cosmological parameters and working in the opposite direction. In particular, the abundance of dark matter halos at the lens redshift can be observationally constrained within the halo model of structure formation. In \cite{CUE21}, the authors modeled the halo mass function according to the popular Sheth \& Tormen \citep{ST99} and Tinker \citep{tinker08} analytical expressions and aimed to constrain its parameters along with the HOD of GAMA lenses in the non-tomographic approach. They found that, while both options described the data decently, the three-parameter Sheth \& Tormen model performed better since the results did not strongly depend on the choice of priors. Contrary to typical values found in cosmological simulations, they also concluded that the $p$ parameter in the model preferred negative values and raised the point of the allowed parameter range in halo mass function fits. Although their results are in agreement with traditional values within the uncertainties, the authors hinted at a slightly higher number of dark matter halos hosting massive galaxies and small galaxy groups.

\begin{figure}[h]
    \centering
    \includegraphics[width=0.85\linewidth]{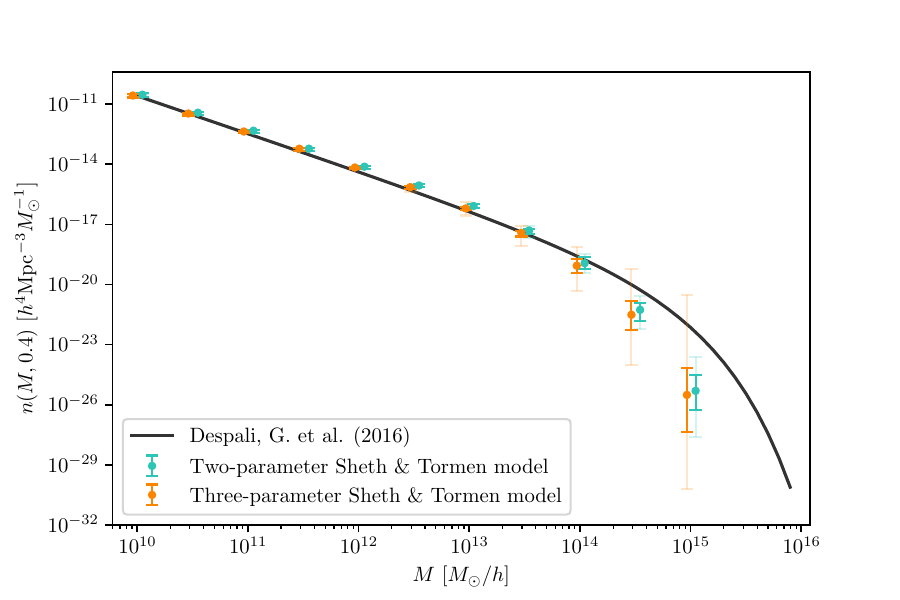}
    \caption{Halo mass function predictions from the two- (teal) and three-parameter (orange) Sheth \& Tormen models using the GAMA/H-ATLAS cross-correlation in the tomographic setup \citep{CUE22}. Bold and faint errorbars depict 68\% and 95\% credibility, respectively. The Sheth \& Tormen best fit by \cite{despali16} using numerical simulations is shown as the black solid line. This figure has been modified with respect to the original source by \cite{CUE22}.} 
    \label{hmf}
\end{figure}

The analysis was later performed by \cite{CUE22} in a tomographic setup. Under the assumption of halo mass function universality, the uncertainties on the three parameters of the Sheth \& Tormen model were significantly reduced with respect to \cite{CUE21}, yielding a 2$\sigma$ prediction of a higher number of dark matter halos of masses between $\sim10^{10}$ and $\sim10^{12}$ $M_{\odot}/h$, as shown in orange in Figure \ref{hmf}. Imposing that all mass in the Universe be bound up into dark matter halos within the halo model formalism effectively normalizes the halo mass function and translates into one less parameter for the model. In that case, the authors found a much more stringent result, predicting more halos of masses $\lesssim 10^{13}M_{\odot}/h$ and fewer halos of masses $\gtrsim10^{14}M_{\odot}/h$ at more than $3\sigma$ with respect to typical cosmological simulations. The possibility of a violation of redshift universality was also tested in \cite{CUE22} by modeling both the HOD and the halo mass function in each redshift bin separately. However, due to the large uncertainties in this case, the hypothesis of universality could not be rejected.

A different avenue that has been explored involves measuring magnification bias on \emph{Herschel} SMGs down from several hundreds kpc by building a stacked map of background sources around the lenses. This approach was first introduced by \cite{BON19}, where QSOs at $z\sim0.7$ acting as gravitational lenses were stacked. A NFW fit to the data yielded a cluster-sized halo mass of $M_{200c}\sim10^{14}M_{\odot}/h$ and a concentration of $c=3.5^{+0.5}_{-0.3}$, in agreement with the literature. This idea was further explored by \cite{fer22}, who measured magnification bias on high-redshift SMGs via stacked maps of galaxy clusters, finding that no individual profile seemed to be able to jointly fit all scales. A similar finding is reported by \cite{CRE22} for lens galaxies and QSOs, for which a double NFW profile is needed. The situation reflects a difference between the inner and outer parts of the lens (separated at 10 arcsec), with halo masses of $\sim 10^{13.3}M_{\odot}/h$ and $\sim 10^{12.8}M{\odot}/h$ in the case of galaxy lenses, signaling isolated galaxies or central galaxies of a small group of dwarfs. QSOs, while featuring similar halo masses ($\sim 10^{13.5}M_{\odot}/h$) in both regimes, were found to have substantially different values for their NFW concentration. Lastly, taking advantage of the better angular resolution of WISE, \cite{CRE24} performed a similar analysis using all three kinds of lens samples, but with a cross-matched catalog between the WISE and H-ATLAS positions of high-redshift SMGs. The improvement in resolution with respect to previous works revealed two distinct regimes: a central excess region and an outer power-law profile, separated by a lack of signal around $\theta\sim10$ arcsec seemingly sourced by (strong) lensing effects.

\section{Outlook and challenges}
\label{sec5}

Ever since its first robust detection in the early 2000s, cosmic magnification has languished in the shadow of shear-based cosmological probes and most efforts have been devoted to correcting it as a bias rather than leveraging it as an observable in itself. As a consequence, its scientific development has been modest, especially using high-redshift SMGs as the background sample. 
However, this can be regarded as an opportunity to progress in a largely unexplored domain, with possibilities from the point of view of data, simulations and theory.

Cosmic magnification on background SMGs has relied so far on the robust and uniform detection of these sources in a large portion of the sky. As the widest-area submillimeter survey, H-ATLAS has been the only SMG background catalog used for this purpose, with the only exception of the HerMES survey in \cite{WANG11}. Along this line, the search for a larger and homogeneous sample of high-redshift SMGs in other areas surveyed by \emph{Herschel} should be prioritized, given the obvious downside of the absence of a wider-area submillimeter survey in the foreseeable future. It should be emphasized that such a program would be of great use to the astrophysical community, not only to maximize the potential of cosmic magnification with a much larger background sample but also to better characterize dust-obscured star formation and the evolution of the progenitors of massive local elliptical galaxies. Regardless of this, the twenty-year gap in more general cosmic magnificaiton analyses (for instance, using background QSOs) can be seized as a promising opportunity to leverage updated archival data of these sources for cosmological goals.

Regarding the lenses, the use of a sample of photometrically detected galaxies with reliable redshifts should be studied to assess the potential improvement from a more numerous foreground catalog. While the spectroscopic GAMA survey was coordinated with H-ATLAS and enabled a clean tomographic division of the lenses, a definite analysis regarding its limitations as a foreground sample has yet to be carried out. In addition, whereas all cosmological results up to date have relied on cosmic magnification alone, a joint fit with the clustering of foreground galaxies would yield tighter constraints and exploit the potential of a 2x2pt analysis as it is common practice in shear galaxy-galaxy lensing. While this goal was already pursued in \cite{CUE24a}, an unexplained inconsistency was found in the simultaneous modeling of both observables. A thorough analysis is needed to probe the issue in depth and determine whether it reflects a fundamental problem or instead arises from poor modeling or unaccounted-for measurement systematics. 

Along this line, cosmological simulations can be leveraged to assess the validity of both the theoretical models and the internal covariance estimations used for real data. In particular, one can set a lower limit for the scales over which a linear galaxy bias or a halo model of structure formation are valid as well as analyze whether a covariance based on Bootstrap or Jackknife algorithms correctly reflects the correlation between different scales. Although such studies are needed for the maturity of the field in itself, they can also answer some of the currently unexplained questions regarding sampling variance or the impact of observational systematics via their injection into raw mock data.

%%%%%%%%%%%%%%%%%%%%%%%%%%%%%%%%%%%%%%%%%%
\section{Conclusions}
\label{sec6}

This review has summarized the state of the art of cosmic magnification on background SMGs, with a description of the theoretical basis, the standard methodology of the field and the latest cosmological results. Cosmic magnification has been overshadowed by shear-based observables on the grounds of low statistical significance, despite its equally fundamental nature and its independence and complementarity as a cosmological probe. However, the provision of a large number of high-redshift SMGs from the \emph{Herschel} space observatory changed the outlook of cosmic magnification, since these sources constitute the optimal background sample for such studies thanks to their high redshift, faint optical emission and steep number counts. 

Most analyses on cosmic magnification with background SMGs have focused on deriving cosmological constraints from the cross-correlation between the GAMA and H-ATLAS surveys, with a number of methodological studies and refinements to the measurement strategy and data treatment. While limited by the much smaller area compared to cosmic shear or galaxy-galaxy lensing works, the results in terms of constraining power have been remarkable. Additionally, a number of points have been raised regarding the importance of calibrating the slope of the SMG number counts or the effect of sampling variance, which need to be further addressed for a higher robustness. 

However, the field has not received much attention and, as a consequence, it evolves slowly (but steadily). Fortunately, this opens a wealth of possibilities to test from the point of view of theory, simulations and observations. Indeed, the inclusion of more complex models for the non-linear power spectrum is needed to account for non-linear and baryonic effects, via augmented halo models including non-linear halo bias or simulation-based emulators. In addition, cosmological simulations can be leveraged to predict the signal, test theoretical models and estimate the covariance matrix externally. From an observational perspective, the necessity for a wider-area far-infrared/submillimeter survey is clear, since it would provide larger uniform samples of high-redshift SMGs that would prove extremely useful for both the astrophysical and cosmological communities. Although it dates back decades, cosmic magnification is still an underexplored field brimming with potential to become a competitive cosmological probe.

%%%%%%%%%%%%%%%%%%%%%%%%%%%%%%%%%%%%%%%%%%
%%%%%%%%%%%%%%%%%%%%%%%%%%%%%%%%%%%%%%%%%%
\vspace{6pt} 

%%%%%%%%%%%%%%%%%%%%%%%%%%%%%%%%%%%%%%%%%%
%% optional
%\supplementary{The following supporting information can be downloaded at:  \linksupplementary{s1}, Figure S1: title; Table S1: title; Video S1: title.}


% Only used for preprtints:
% \supplementary{The following supporting information can be downloaded at the website of this paper posted on \href{https://www.preprints.org/}{Preprints.org}.}

%\begin{tabularx}{\textwidth}{lll}
%\toprule
%\textbf{Name} & \textbf{Type} & \textbf{Description} \\
%\midrule
%S1 & Python script (.py) & Script of python source code used in XX \\
%S2 & Text (.txt) & Script of modelling code used to make Figure X \\
%S3 & Text (.txt) & Raw data from experiment X \\
%S4 & Video (.mp4) & Video demonstrating the hardware in use \\
%\bottomrule
%\end{tabularx}
%}

%%%%%%%%%%%%%%%%%%%%%%%%%%%%%%%%%%%%%%%%%%
\authorcontributions{Writing---original draft preparation, M.C.; writing---review and editing, M.C., J.G., L.B. and A.L. All authors have read and agreed to the published version of the manuscript.}

\funding{This research received no external funding.}

\dataavailability{No new data were created or analyzed in this study. Data sharing is
not applicable to this article.}

\conflictsofinterest{The authors declare no conflicts of interest.} 

%%%%%%%%%%%%%%%%%%%%%%%%%%%%%%%%%%%%%%%%%%
%% Optional

%% Only for journal Encyclopedia
%\entrylink{The Link to this entry published on the encyclopedia platform.}

%%%%%%%%%%%%%%%%%%%%%%%%%%%%%%%%%%%%%%%%%%
%\isPreprints{}{% This command is only used for ``preprints''.
\begin{adjustwidth}{-\extralength}{0cm}

%\reftitle{References}


\bibliography{main.bib}

%=====================================
% References, variant B: internal bibliography
%=====================================

% Chicago format (Used for journal: arts, genealogy, histories, humanities, jintelligence, laws, literature, religions, risks, socsci)
\isChicagoStyle{%

}{}

% APA format (Used for journal: admsci, behavsci, businesses, econometrics, economies, education, ejihpe, games, humans, ijfs, journalmedia, jrfm, languages, psycholint, publications, tourismhosp, youth)
\isAPAStyle{%

}{}

% If authors have biography, please use the format below
%\section*{Short Biography of Authors}
%\bio
%{\raisebox{-0.35cm}{\includegraphics[width=3.5cm,height=5.3cm,clip,keepaspectratio]{Definitions/author1.pdf}}}
%{\textbf{Firstname Lastname} Biography of first author}
%
%\bio
%{\raisebox{-0.35cm}{\includegraphics[width=3.5cm,height=5.3cm,clip,keepaspectratio]{Definitions/author2.jpg}}}
%{\textbf{Firstname Lastname} Biography of second author}

% For the MDPI journals use author-date citation, please follow the formatting guidelines on http://www.mdpi.com/authors/references
% To cite two works by the same author: \citeauthor{ref-journal-1a} (\citeyear{ref-journal-1a}, \citeyear{ref-journal-1b}). This produces: Whittaker (1967, 1975)
% To cite two works by the same author with specific pages: \citeauthor{ref-journal-3a} (\citeyear{ref-journal-3a}, p. 328; \citeyear{ref-journal-3b}, p.475). This produces: Wong (1999, p. 328; 2000, p. 475)

%%%%%%%%%%%%%%%%%%%%%%%%%%%%%%%%%%%%%%%%%%
%% for journal Sci
%\reviewreports{\\
%Reviewer 1 comments and authors’ response\\
%Reviewer 2 comments and authors’ response\\
%Reviewer 3 comments and authors’ response
%}
%%%%%%%%%%%%%%%%%%%%%%%%%%%%%%%%%%%%%%%%%%
\PublishersNote{}
%\isPreprints{}{% This command is only used for ``preprints''.
\end{adjustwidth}
\end{document}